\documentclass[twocolumn,superscriptaddress,amsmath,amssymb]{revtex4}
\usepackage[utf8]{inputenc}
\usepackage[draft]{fixme}
\usepackage[ruled]{algorithm}
\usepackage{algpseudocode}
\usepackage{graphicx,xspace}
\algnotext{EndFor}
\algnotext{EndIf}

\newcommand{\ket}[1]{| #1 \rangle}

\newcommand{\Class}{\mathcal{C}}

\begin{document}
\title{Algorithmic pseudorandomness in quantum setups}

\author{Ariel Bendersky}
\affiliation{ICFO-Institut de Ciencies Fotoniques, Mediterranean Technology Park,
08860 Castelldefels (Barcelona), Spain}
\author{Gonzalo de la Torre}
\affiliation{ICFO-Institut de Ciencies Fotoniques, Mediterranean Technology Park,
08860 Castelldefels (Barcelona), Spain}
\author{Gabriel Senno}
\affiliation{Departamento de Computaci\'on, FCEN, Universidad de Buenos Aires, Buenos Aires, Argentina.}
\author{Santiago Figueira}
\affiliation{Departamento de Computaci\'on, FCEN, Universidad de Buenos Aires, Buenos Aires, Argentina.}
\affiliation{CONICET, Argentina}
\author{Antonio Acin}
\affiliation{ICFO-Institut de Ciencies Fotoniques, Mediterranean Technology Park,
08860 Castelldefels (Barcelona), Spain}

\begin{abstract}

The Church-Turing thesis is one of the pillars of computer science; it postulates that every classical system has equivalent computability power to the so-called Turing machine. While this thesis is crucial for our understanding of computing devices, its implications in other scientific fields have hardly been explored.
Here we start this research programme in the context of quantum physics and show that computer science laws have profound implications for some of the most fundamental results of the theory. We first show how they question our knowledge on what a mixed quantum state is, as we identify situations in which ensembles of quantum states defining the same mixed state, indistinguishable according to the quantum postulates, do become distinguishable when prepared by a computer.
We also show a new loophole for Bell-like experiments: if some of the parties in a Bell-like experiment use a computer to decide which measurements to make, then the computational resources of an eavesdropper have to be limited in order to have a proper observation of non-locality. Our work opens a new direction in the search for a framework unifying computer science and quantum physics.

\end{abstract}

\maketitle

Quantum theory stands as one of the most successful and experimentally confirmed theories to date, with not a single experiment shown to be in disagreement. Its foundations, intensely debated in the early days of the theory~\cite{Bohr1935,Einstein1935}, remain an active area of research with many recent insightful results, such as the possibility to derive quantum theory from physical axioms~\cite{Masanes2011Derivation}, proving the completeness of quantum theory~\cite{Colbeck2011a}, or establishing that quantum states are real~\cite{Pusey2012Reality}.

On the other hand, computability theory studies which functions can be calculated by algorithmic means and which cannot, and, more generally, their degree of uncomputability. The field emerged in the 1930s with the independent works of Alan Turing~\cite{Turing1936} and Alonzo Church~\cite{Church1936} who introduced two equivalent formalizations of the intuitive concept of \emph{algorithm}. The fact that both models of computation turned out to be equivalent lead Stephen Kleene~\cite{Kleene1967} to postulate what is now known as the Church-Turing thesis:

\begin{quote}
\sl any function `naturally to be regarded as computable' (i.e.\ calculable by algorithmic means) is computable by the formal model of Turing machines.
\end{quote}
This thesis has been greatly strengthened by the fact that all the formal models of computation defined so far 
have been shown to be at most as expressive as the classical Turing machines in terms of the class of functions they compute.

Until now, the relationship between quantum mechanics and the Church-Turing thesis has been concerned on how the first one can affect the latter (see, for instance, \cite{Arrighi2012}). In this article, however, we introduce an opposed and new research program: implications of computer science principles for quantum physics.

As the first steps in that direction, we present two results. First we show that computers impose a limitation when it comes to producing a mixed state as a classical mixture of pure quantum states. It turns
out that with the sole knowledge that the classical mixture is performed by a computer, situations that seem not to be distinguishable turn out to be so. This has direct implications since mixed states are prepared this way in many experiments\cite{amselem2009experimental, PhysRevLett.105.130501}. Secondly, when it comes to Bell-like experiments to test non-locality, another distinctive feature of quantum mechanics, we show that if the measurement independence between the two parties \cite{Conway2006,Koh2012,Hall2010,Barrett2010} is achieved via private computable pseudo random number generators, an eavesdropper can start guessing their inputs from the information on their previous inputs, thus leading to a new computability loophole for Bell tests.

Formally, our results apply only to computers. This has already considerable practical consequences, since almost every experimental setup is controlled by computers. However, one can argue that, due to the widely accepted physical interpretation of the Church-Turing thesis:
\begin{quote}
\sl the behaviour of any discrete physical system evolving according to the laws of classical mechanics is computable by a Turing machine,
\end{quote}
our results, in fact, apply to every classical system and hence, the limitations that we show are fundamental.

\section{Proper mixed state preparation and the Church-Turing thesis}
\newcommand{\tup}[1]{\langle#1\rangle}
\newcommand{\cantor}{2^\omega}
\newcommand{\words}{2^*}
\newcommand{\uph}{\upharpoonright}
\newcommand{\NN}{\mathbb{N}}


We start by considering one of the basic parts of quantum theory: the concept of a mixed state \cite{Nielsen2000}. We will see that although well understood by physicists, their nature and origin can lead to apperent paradoxes when confronted against common computer science tenets.

\begin{figure}[h]
\includegraphics[width=\columnwidth]{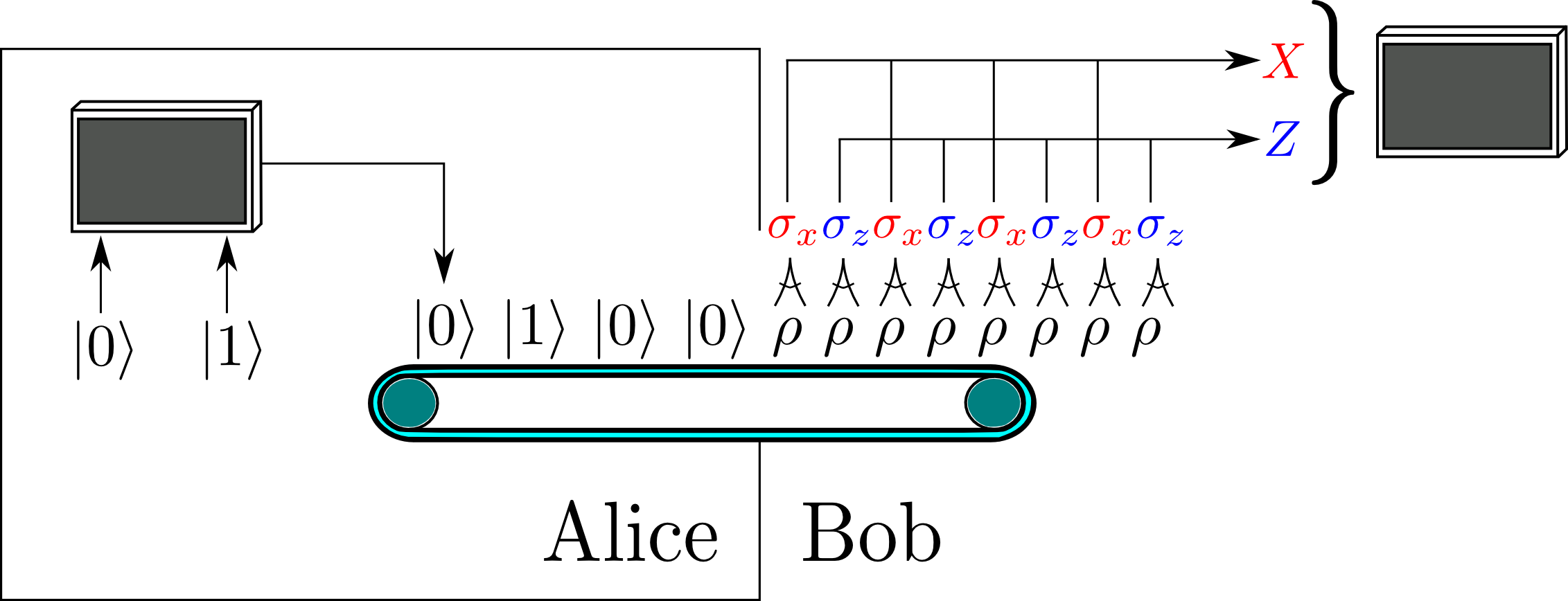}
\caption{Scheme for the preparation of mixed states via classical mixing. Alice uses a computer to choose between $\ket{0}$ and $\ket{1}$ (or $\ket{+}$ and $\ket{-}$). From Bob's perspective, since he doesn't know the computer program Alice is using, it's a mixed state $\rho$. To distinguish both possible preparations he measures alterantively $\sigma_x$ and $\sigma_z$ and sends the resulting sequences to a computer that will be able to tell which of the sequences is computable (corresponding to the basis in which Alice prepared the state) and which is a fair coin tossing (corresponding to the other basis).\label{figMixedState}}
\end{figure}

Let us present now the following preparation of a mixed state: A classical computer in an unknown configuration and with unbounded memory is running an unknown and presumably very convoluted algorithm to prepare a mixed state. We have the promise that the computer, understood as a black box, is mixing evenly either the single qubit eigenstates of $\sigma_z$  $\{ \ket{0}, \ket{1} \}$ or the states $\{ \ket{+}, \ket{-}\}$, where $\ket{\pm}=\frac{1}{\sqrt{2}}\left(\ket{0}\pm\ket{1}\right)$ as seen in Fig. \ref{figMixedState}.     Is there any operational procedure to decide which of the two ensembles are being mixed for an experimenter (Bob) who cannot open the black box? Even though one would be tempted to assign to both preparations the identity state $\rho=\frac{\mathbb I}{2}$, our results show that the fact that the mixing procedure was performed in a computable way leaves a trace which allows us to distinguish both mixtures in finite time and with arbitrarily high success probability. It is worth
mentioning that having a computer mixing the state doesn't imply that the sequence in which it mixes the state is periodic or anything. 
In fact, there exist normal sequences (e.g. those which satisfy the law of large numbers in a generalized way), or other even `more random' sequences which are computable in polynomial time \cite{FN13}.

In order to solve this problem, 
Bob measures every qubit that comes out of the black box on an odd position in the basis of eigenstates of $\sigma_z$, yielding a binary sequence of measurement results $Z$. He also measures every qubit on an even position in the basis of eigenstates of $\sigma_x$ obtaining a binary sequence $X$. This way Bob obtains two binary sequences, as can be seen in Fig. \ref{figMixedState}. The one corresponding to the choice of measurement that matches the preparation basis is computable, and the other one corresponds to a fair coin tossing. Therefore we need an algorithm that given two sequences, one computable and one arising from a fair coin tossing, is able to tell us which is which. We will show now one such algorithm, that can perform this task in finite time and with an arbitrarily high probability of success.

\begin{figure}[th]
\includegraphics[width=\columnwidth]{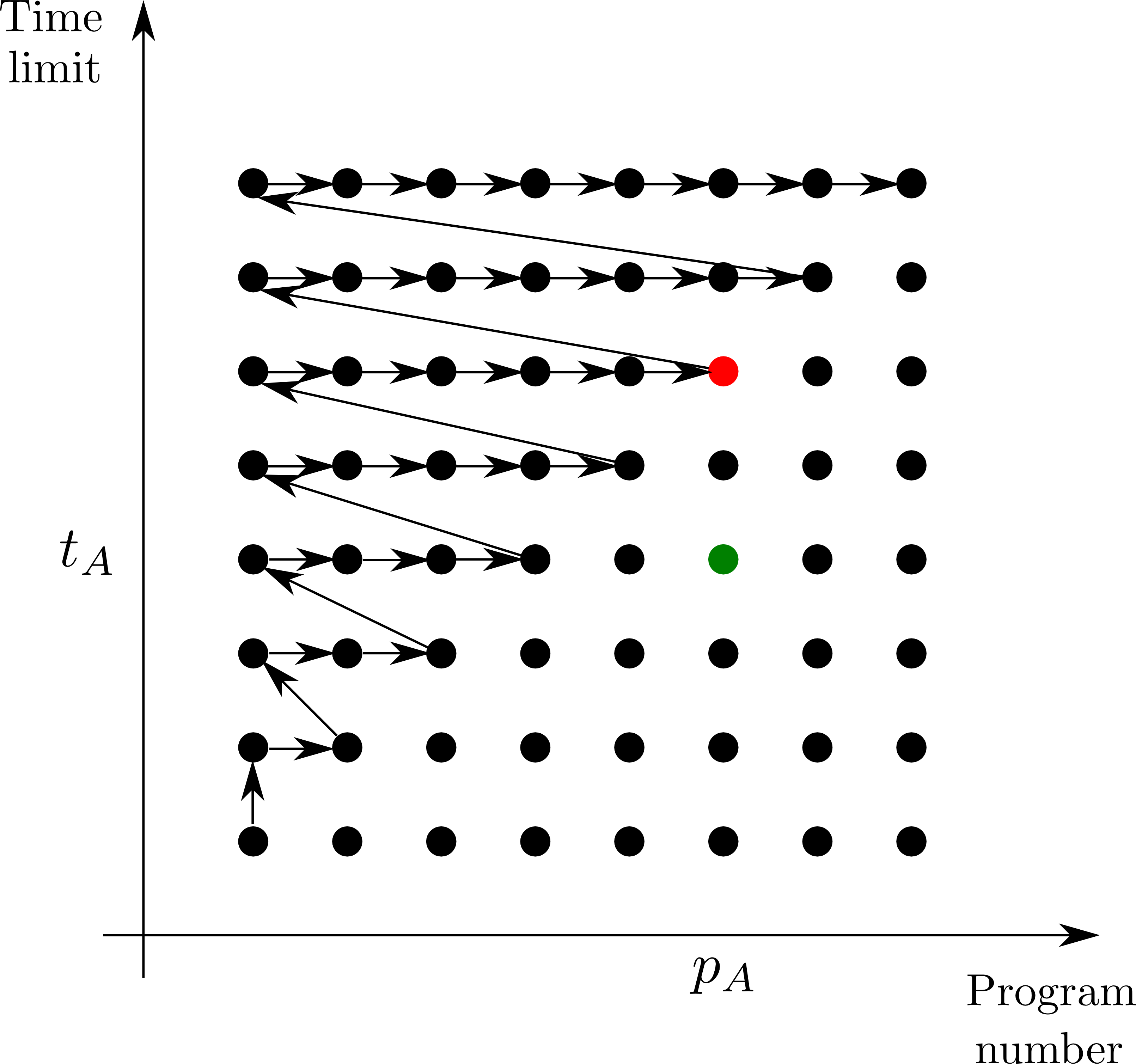}
\caption{To decide which of the two sequences is computable, we dovetail between program number and timesteps that each program is allowed to run. On green we show the actual program  $p_A$ that was used to prepare the state, and the number of timestpes $t_A$ it takes to generate a $k|p_A|$ long prefix. The latest halting condition for our algorithm is shown in red, although it might halt before that with either a wrong recognition or a correct one.\label{figDT}}
\end{figure}

To distinguish which of the two sequences is computable we dovetail between program number (the programs are computably enumerable) and maximum time steps that we allow each program to run (that is, we run program 1 for 1 timestep, then programs 1 and 2 for 2 timesteps and so on), as is a common technique in computability theory. For each program $p$ of length $|p|$ we will compare the first $k|p|$ output bits with the corresponding prefixes of both sequences, where $k$ is an integer constant depending on
the probability of success we are looking for. Whenever we find a match for the first $k|p|$ bits, we halt. Fig. \ref{figDT} depicts the dovetailing algorithm.
It is straightforward to see that this algorithm always gives an answer, and the probability of making a mistake is less than $O(2^{-k})$. Therefore we can guess in finite time and with an arbitrarily high probability of success (by setting $k$ we adjust the probability of success).

The complete algorithm to distinguish a fair coin from a computable sequence is Algorithm \ref{alg:distinguish} below, where $X\uph k|p|$ denotes the fist $k|p|$ bits of the sequence $X$. 

\begin{algorithm}\caption{The distinguishing protocol. $U_t(p)$ is a universal Turing machine that runs program $p$ for $t$ timesteps. The two `for' loops correspond to the dovetailing.}\label{alg:distinguish}
\begin{algorithmic}
\Require $k \in \NN$ and $X,Z\in\cantor$, two bit sequences with the promise that one of them is computable and other is not.
\Ensure `$X$' or `$Z$' as the candidate for being computable; wrong answer with probability bounded by $O(2^{-k})$.
\For{$t=0,1,2\dots$}
    \For{$p=0,\dots,t$}
            \If {$U_t(p)=X\uph k|p|$} \State output `$X$' and halt \EndIf
            \If {$U_t(p)=Z\uph k|p|$} \State output `$Z$' and halt \EndIf
    \EndFor
\EndFor
\end{algorithmic}
\end{algorithm}

Note  that, at a given iteration of the algorithm, it may perfectly be the case that program $p$ has not been able to produce in $t$ time steps the $k|p|$ symbols needed to check the halting condition. If this is the case, the algorithm simply keeps running and moves to the next program. However, the algorithm will for sure halt as it will run the actual program used in the blackbox at some finite time. For a detailed explanation on how Algorithm \ref{alg:distinguish} works and its probability of success, see Appendix \ref{ApDist}.

It is an interesting open question to study the effect of noise in the previous algorithm. As a first step, we have considered a rather simple noise model in the state preparations and measurements described by a flip probability in the observed symbols $r$. That is, we consider the situation in which those results obtained when measuring the quantum states in the actual basis used by the box are correct with probability $1-r$ (this simple noise has no effect on the results of measurements performed in the wrong basis). As shown in Appendix \ref{ApDist}, there is another slightly more complex algorithm that still halts with arbitrarily small error probaility whenever $r\lesssim 0.21$.

The previous algorithm is of course very demanding, but proves that ensembles of states defining the same mixed state are in principle distinguishable when prepared by classical computing devices. This questions our understanding of mixed quantum states and leaves quantum mixtures (either by using a part of a larger etangled system or a quantum random number generator) as the only way to create them. 

\section{Bell inequality loophole}
\newcommand{\CC}{\mathcal{C}}


Non-locality is another of the most intrinsic features of quantum mechanics ~\cite{Einstein1935, Bell1964, Bell1966}. The standard Bell scenario is described by two distant observers who can perform $m$ possible measurements of $d$ possible outputs on some given devices. The measurements are arranged so that they define space-like separated events. 
It is convenient for what follows to rephrase the standard Bell scenario in cryptographic terms, as in \cite{PhysRevA.66.042111, pironio2010random,PhysRevA.87.012336}. In this approach, Alice and Bob get the devices from a non-trusted provider Eve. The standard local EPR models correspond to classical preparations in which the devices generate the measurement results given the choice of measurements, but independently of the input chosen by the other party. Bell inequalities are conditions satisfied by all these preparations, even when having access to all the measurement choices and results produced in previous steps \cite{PhysRevA.66.042111}. 
In turn, quantum correlations, obtained for example by measuring a maximally entangled two-qubit state with non-commuting measurements, can violate these inequalities. The violation of a Bell inequality witnesses the existence of non-local correlations and can be used by Alice and Bob to certify the quantum nature of their devices.

\begin{figure}[h]
\includegraphics[width=\columnwidth]{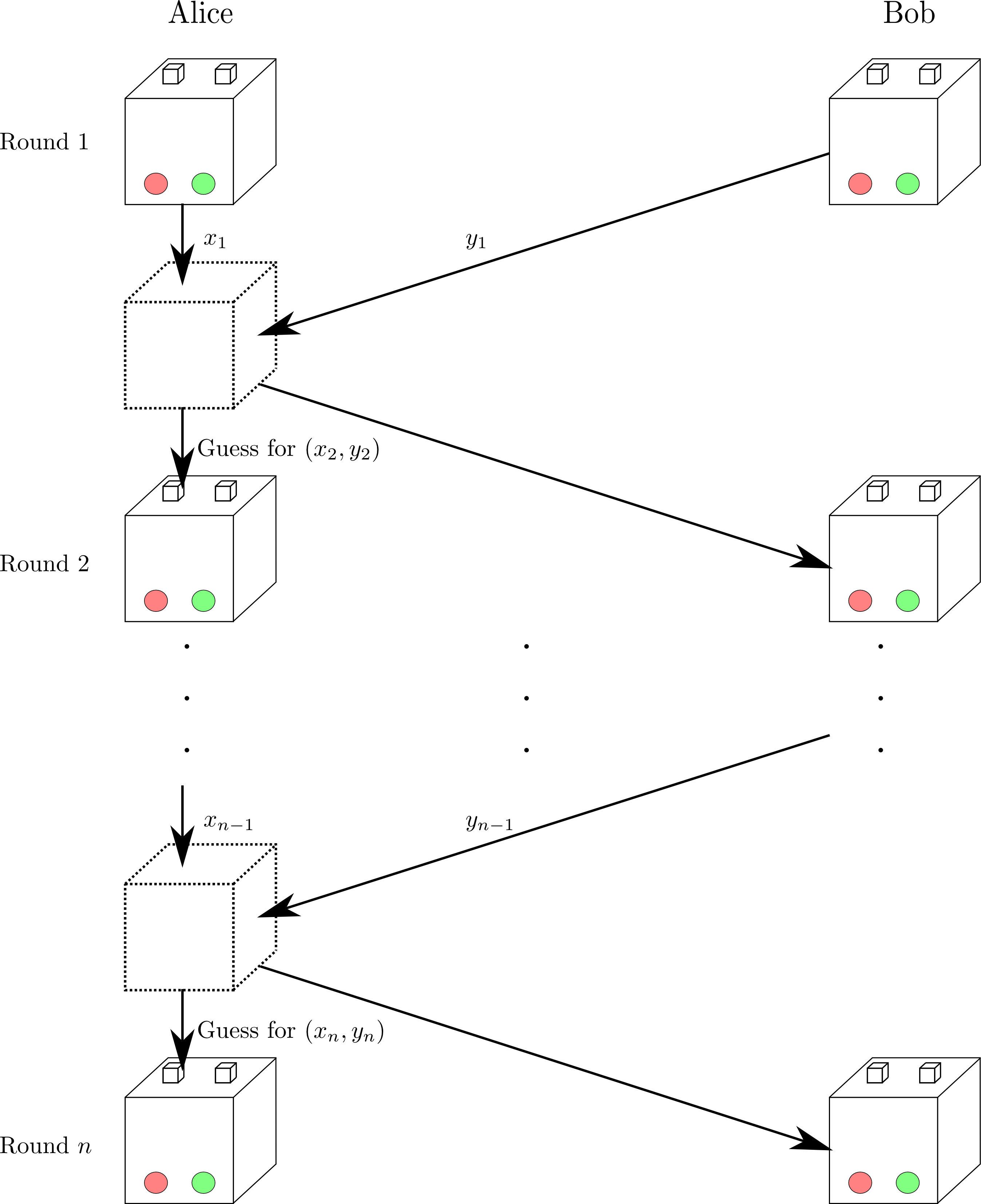}
\caption{ Scheme for the Bell inequality computability loophole. After each round $i$, Alice's box receives the information about  Bob's choice of measurement $y_i$. Using the information from all previous choices of inputs for both parties, Alice's box makes a prediction on what the next round inputs will be by using the presented algorithm. For simplicity in the notation, the attack is shown for a Bell test involving two measurements per party. As long as Alice, Bob or both use a computable sequence for their choice, the guess will start being correct after a number of rounds. Once this happens, the boxes can simulate any probability distribution, both local and non-local. Therefore Alice and Bob can not rule out an eavesdropper having prepared their boxes. \label{figAliceBob}}
\end{figure}

In what follows, it is shown how a classical Eve can mimick a Bell inequality violation when the measurement choices on Alice and Bob are performed following an algorithm, which is a standard practice in many Bell experiments to date. As above, it is not assumed that the algorithm is known by the eavesdropper. The result can be seen as a new loophole, named the \emph{computability loophole}. For this loophole to apply, Eve has to make use of the inputs and outputs produced by the parties in previous steps  \cite{PhysRevA.66.042111, pironio2010random,PhysRevA.87.012336}, as shown in Fig.~\ref{figAliceBob}.


\begin{figure}[h]
\includegraphics[width=\columnwidth]{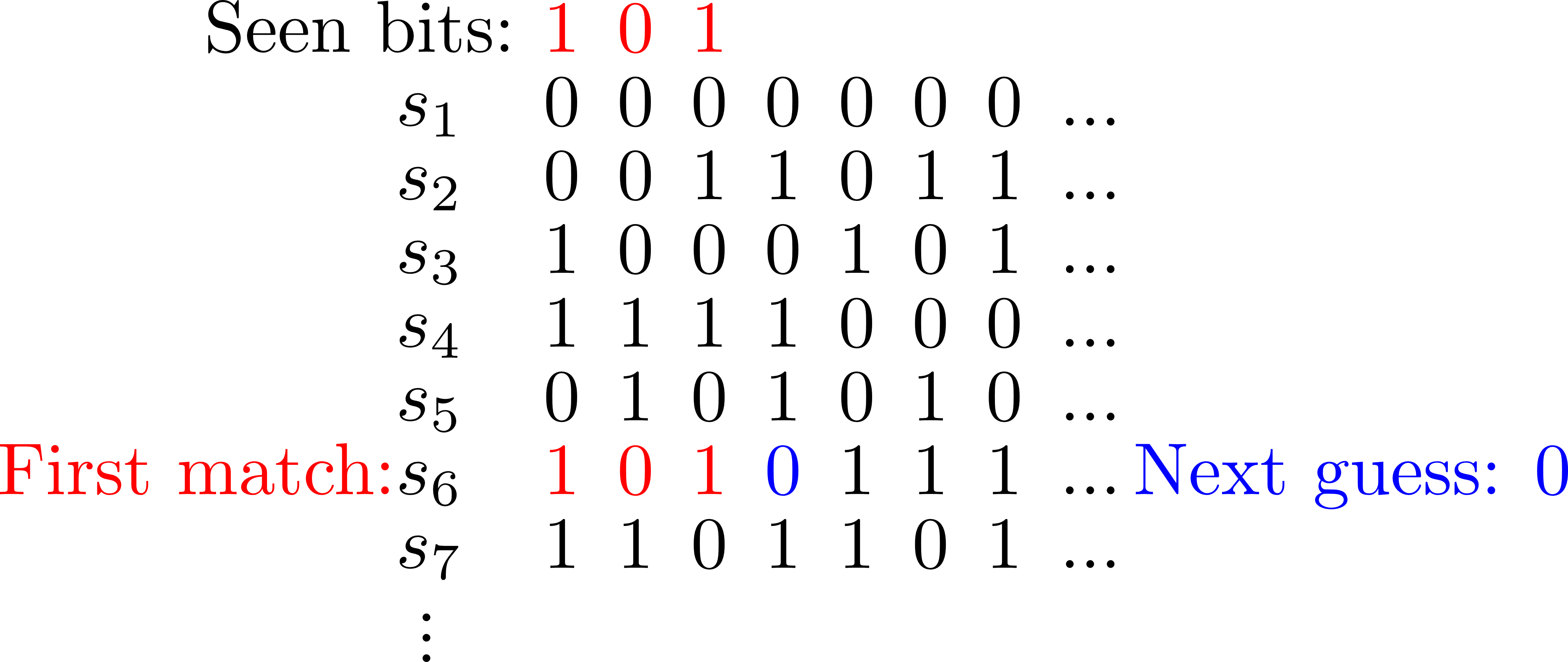}
\caption{Algorithm to predict bits from previously seen ones. For simplicity it is presented for the case of $2$ inputs. Alice's box enumerates all sequences $s_i$ from the complexity class $\CC$ and picks the first one whose first bits coincide with the ones seen. The guess Eve will make is the next bit from that sequence. If the actual sequence one wants to predict belongs to class $\CC$, at some point, after seeing enough bits, all the predictions will be correct.\label{figLearn}}
\end{figure}

The computability loophole is rather simple and works as follows: one of the devices, say Alice's, uses all the inputs chosen in previous steps by both parties to guess the next ones. For that, a time complexity class $\CC$ is initially chosen by Eve. Since algorithms in such a class are computably enumerable, Alice's device will check all of them until it finds one that matches Alice's (or Bob's) bits given so far. A guess for the next inputs is done based on that algorithm (see Fig. \ref{figLearn} for a representation of the algorithm) and communicated to the other device. If Eve's class $\CC$ includes Alice's and/or Bob's algorithms, at some point the device will start guessing correctly and will keep doing so forever. Of course, once the devices are able to guess the inputs of at least one of the parties, they can easily produce non-local correlations.

It should be noticed that since Alice's and Bob's algorithms belong to some time complexity class, they can never rule out such an eavesdropper. On the other hand, the eavesdropper, when choosing the class $\CC$, is imposing how hard it is for Alice and Bob to avoid the loophole. The
algorithm is shown as Algorithm \ref{alg:nextvalue}, and for a more detailed description the reader is refered to Appendix \ref{apBell}.

\begin{algorithm}\caption{A next-value algorithm for a time class $\Class$ with bound $t$}\label{alg:nextvalue}
\begin{algorithmic}
    \Require $n \in \NN$
    \Ensure $g(n)$, the next-value function for $\Class$.
    \State Let $M_e$ be an enumeration of all Turing machines.
    \State Let $n=\langle m_0,\dots, m_{n-1}\rangle$ be the already seen bits from the sequence.
    \State Let $\langle e,c\rangle$ be the least number such that
        \begin{enumerate}
        \item[i.] for $i\in\{0,\dots,n\}$, $M_e(i)$ halts after at most $c\cdot t(|i|)$ many steps, where $M_e$ evaluates function number $e$ from $\Class$ and $t$ is the computable time bound for class $\Class$.
        \item[ii.] for $i\in\{0,\dots,n-1\}$ $M_e(n)$ outputs $m_i$
        \end{enumerate}
    \State Output $M_e(n)$
\end{algorithmic}
\end{algorithm}

Apart from being very demanding, a criticism to this loophole is that it only works in the long run, meaning that Alice and Bob will not see a fake violation of a Bell inequality unless they run their experiment for long enough. But this brings the question of what's the validity of a violation that, in the long run, would have admited a local model. The only way to escape this loophole is by using a quantum random generator for the inputs, however, it is highly undesirable to depend on a non-local theory to test non-locality.

\section{Discussion}

The Church--Turing thesis is one of the most accepted postulates from computer science. As such, one can wonder what consequences would it have if it were, indeed, a law of Nature. We started here this research program in the context of quantum physics by showing how it questions the understanding of what a proper mixed quantum state is and introducing a new loophole for Bell tests. The study of these questions is essential in any attempt to create a unifying theory merging information laws and quantum physics.


\section{Acknowledgements}

This work was supported by the ERC CoG QITBOX, the John Templeton Foundation, the Argentinian UBA (UBACyT 20020110100025) and ANPCyT (PICT-2011-0365). GdlT acknowledges support from Spanish FPI grant (FIS2010-14830).

\appendix

\section{Distinguishing two computable preparations of the same mixed quantum state\label{ApDist}}

In this section we discuss with details the protocol to
distinguish two ensembles of pure states that apparently yield the
same density matrix, given the assumption that the mixture has
been prepared in a computable way. To do so, we will first reduce
this scenario to a different problem in classical information
theory relating infinite binary sequences, which we show to be
solvable. Then we see how this second problem can be solved in
finite time with arbitrarily small error probability. Finally, we
present a slight modification of the algorithm that makes it
robust against a simple noise model.

\subsection{The distinction protocol}

The distinguishability scenario we are interested in is as
follows: Alice is presented with two bags of quantum systems, one
having systems in the state $\ket{0}$ and the other having systems
in the state $\ket{1}$. At random, she chooses one bag at a time
and sends a state from that bag to Bob. Bob's state is
$\rho={\mathbb I}/2$, as he has no information on the prepared
state. Now imagine the same scenario, but instead of those two
bags Alice has one with the state $\ket{+}$ and another with the
state $\ket{-}$, and she proceeds in the same way. It is clear
that these two situations are indistinguishable from Bob's point
of view, as they are supposed to define the same mixed quantum
state.

Consider the same two situations, but now, in order to choose from
which bag to pick the state, Alice uses the bits of a computable
binary sequence, that is, a binary sequence produced by an
algorithm. Ideally such algorithm is a pseudo-random number
generator, whose output `looks like' typical coin tossing (i.e.\
satisfying for instance the law of large numbers, as well as any
other reasonable law of randomness~\cite{Li2008}). We next see that,
if Alice's sequence is computable, Bob can distinguish between
both situations in finite time and with an arbitrarily high
probability of success.

Bob's protocol works as follows: he measures Alice's first qubit
in the basis of eigenstates of $\sigma_x$, the second qubit in the
basis of eigenstates of $\sigma_z$, and so on, measuring every odd
qubit in the $\sigma_x$ basis and every even qubit in the
$\sigma_z$ basis.
The output of the measurements yields in the limit two 
infinite binary sequences: $X$, obtained from the measurements on
the odd qubits and $Z$, obtained from the even qubits. Now $X$ and
$Z$ have a distinctive feature: when Bob measures in the same
basis as Alice prepared the states, the sequence obtained is
computable (because it is either the odd or the even bits of a
computable sequence), and when the measurement is performed in the
other basis the sequence is similar to one obtained from tossing a
fair coin. Thus, if Bob can distinguish a computable sequence from
a fair coin he can tell what was the basis in which Alice prepared
the state. We will see that this is indeed the case with
arbitrarily high probability of success. Both preparations are,
thus, distinguishable. It should be noticed that, since Bob will
only need finite prefixes from both sequences to achieve the
distinction, he just needs to measure a finite number of qubits
received from Alice.

As mentioned, after his measurements, Bob is left with the problem
of distinguishing a computable sequence from a fair coin tossing.
We present an algorithm that can do this with an arbitrarily high
probability of success: given two sequences $X$ and $Z$ and a
desired error probability $e$, our algorithm decides in finite
time which of the two sequences is the computable one, giving a
wrong answer with a probability smaller or equal than $e$. The
hand-wavy idea of the algorithm is to check every program (the set of Turing
machines is enumerable) on an universal Turing machine $U$ until
finding one that reproduces a sufficiently long prefix of either
$X$ or $Z$, say $X$. It is then claimed that $X$ is the computable
sequence, that is, the basis used by Alice for encoding.

\subsection{Distinguishing a fair coin from a computer}\label{sec:classical_result}

\newcommand{\NV}{{\rm NV}}

In this Section we show the algorithm that can distinguish, with
arbitrarily small error probability, a computable sequence from one
arising from a fair coin.

\subsubsection{Background on computability theory}\label{subsec:comp}

Let us first fix some notation. The set of finite strings over the
alphabet $\{0,1\}$ is denoted $\words$ and $\epsilon$ denotes the
empty string. The set of infinite sequences over $\{0,1\}$ is
denoted $\cantor$. If $S\in\cantor$ then $S\uph n$ denotes the
string in $\words$ formed by the first $n$ symbols of $S$. If
$x,y\in\words$ then $x\preceq y$ represents that $x$ is a prefix
of $y$. Any natural number $n$ can be seen as a string in $\words$
via its binary representation.

First we introduce formally the computing model used throughout
this Section. All in all, it is nothing but a particular model
equivalent to a Turing machine, thus having the same computational
power as any computer with unbounded memory. Specifically, we
consider Turing machines $M$ with a reading, a working and an
output tape (the last two being initially blank). The output of
$M$ on input $x\in\words$ is denoted $M(x)\in\words$, and if
$t\geq 0$, $M_t(x)\in\words$ consists of the content of the output
tape in the execution of $M$ on input $x$ by step $t$ ---notice
that this execution needs not be terminal, that is, $M(x)$ needs
not be in a halting state at stage $t$.  A {\em monotone} Turing
machine (see e.g.~\cite[\S2.15]{DHBook}) is a Turing machine whose
output tape is one-way and write-only, meaning that it can append
new bits to the output but it cannot erase previously written
ones. Hence if $M$ is a monotone machine $M_t(x) \preceq M_s(x)$
for $t\leq s$. The computing model of monotone machines is
equivalent to ordinary Turing machines, and for ease of
presentation we work with the former.

A sequence $S\in\cantor$ is {\em computable} if there is a
(monotone) Turing machine $M$ such that for all $n$, $M(n)=S\uph
n$. Equivalently, $S$ is computable if there is a monotone machine
$M$ such that $M(\epsilon)$ ``outputs" $S$, in the sense that
\begin{equation}\label{eqn:inf_comp}
(\forall n)(\exists t)\ M_t(\epsilon)=S\uph n.
\end{equation}
Let $(M_i)_{i\geq 0}$ be an enumeration of all monotone Turing
machines and let $U$ be a monotone Turing machine defined by
$U(\tup{i,x})=M_i(x)$, where $\tup{\cdot,\cdot}:\NN^2\to\NN$ is
any computable pairing function (i.e.\ one that codifies two
numbers in $\NN$ into one and such that both the coding and the
decoding functions are computable). The machine $U$ is universal
for the class of all monotone machines. In other words, $U$ is an
interpreter for the class of all monotone Turing machines, and the
argument $p$ in $U(p)$ is said to be a {\em program} for $U$,
encoding a monotone Turing machine and an input for it.

The notion of computability makes sense when applied to infinite
sequences, as any finite string can be trivially computed by a
very simple (monotone) Turing machine which just hard-codes the
value of the string. Any finite binary string can be extended with
infinitely many symbols in order to obtain either a computable or
an uncomputable sequence. For instance, if $s$ is a finite string
then $s$ followed by a sequence of zeroes is computable; however
$s$ followed by the (binary representation of) the halting
problem~\cite{Turing1936} is not. Since in finite time a Turing
machine can only process finitely many symbols, one cannot decide
in finite time if an infinite sequence is computable or not.

In the following subsection we deal with a related problem:
distinguishing a computable sequence from the output of a `fair
coin' (such as the result of measuring a $\sigma_z$ eigenstate in
the $\sigma_x$ basis, under the assumption that quantum physics is
correct). Notice that since there are countably many computable
sequences, the output of tossing a fair coin gives a
non-computable sequence with probability one. We show that one can
distinguish both cases in finite time and with arbitrarily high
success probability and that this fact has consequences on how
mixed states in quantum mechanics are described.


\subsubsection{The protocol}

As mentioned, the idea of the algorithm is to check every program
until finding one that reproduces a sufficiently long prefix of
either $X$ or $Z$. There are three key points that have to be
taken into account in this idea, namely:

\begin{itemize}
\item It is impossible to know 
if a program halts or not~\cite{Turing1936}. Therefore, checking each single program one after another is not possible.
\item It is still not clear what we mean by \emph{sufficiently long} prefix.
\item We might get a false positive, i.e.\ find a program that reproduces a prefix of the sequence which came form the coin tossing (even if it was not computable).
\end{itemize}

We deal with the first issue by dovetailing between programs and
execution time. Recall that programs for $U$ can be coded by
(binary representations of) natural numbers. The idea of
dovetailing is that we first run program $0$ for $0$ steps, then
we run programs $0$ and $1$ for $1$ step, then we run programs
$0$, $1$ and $2$ for $2$ steps and so on.

To solve the second problem, we check if program $p$ of length
$|p|$ generates (within the time imposed by the dovetailing)
either of the prefixes $X \uph j$ (that is, the first $j$ bits of
$X$) or $Z \uph j$, where $j=k|p|$. That is, every program is
checked against a prefix $k$ times longer than its length. Since a
fair coin generates sequences with mostly non-compressible
prefixes, it most likely will not have a prefix that can be
generated by a $k$ times shorter program, thus allowing us to
detect the computable sequence. And as we will see, the
probability of getting a false positive can be bounded only by a
function of $k$ that goes to $0$ as $k$ goes to infinite, solving
also the third issue.

The pseudo-code for the algorithm that decides which of the sequences is computable will be the following:

\begin{algorithm}\caption{The distinguishing protocol}\label{alg:distinguishAP}
\begin{algorithmic}
\Require $k \in \NN$ and $X,Z\in\cantor$, one of them being computable
\Ensure `$X$' or `$Z$' as the candidate for being computable; wrong answer with probability bounded by $O(2^{-k})$
\For{$t=0,1,2\dots$}
    \For{$p=0,\dots,t$}
            \If {$U_t(p)=X\uph k|p|$} \State output `$X$' and halt \EndIf
            \If {$U_t(p)=Z\uph k|p|$} \State output `$Z$' and halt \EndIf
    \EndFor
\EndFor
\end{algorithmic}
\end{algorithm}
Note that $X$ and $Z$ are infinite sequences, and hence they must
be understood as {\em oracles}~\cite[\S{III}]{S87} in the
effective procedure described above.
Provided that at least one of $X$ or $Z$ is computable, the above
procedure always halt ---and so it only queries finitely many bits
of both $X$ and $Z$. Indeed, in case $S\in\{X,Z\}$ is computable,
there is a monotone Turing machine $M=M_i$ such that $M(\epsilon)$
outputs $S$ in the sense of \eqref{eqn:inf_comp}. Hence for
program $p=\tup{i,\epsilon}$ we have that $U_t(p)=S\uph k|p|$ for
some $t$.

It is important to recall that although both $X$ and $Z$ are infinite sequences, 
we only need to query finite prefixes. From a physical point of
view this means that only finitely many qubits will be needed by
Bob to discover how Alice was preparing the state.


Now, we bound the probability of having a miss-recognition, that
is, the probability $P_{{\rm error}}$ that the above procedure
outputs `$Z$' when $X$ was computable, or viceversa. To do so, we bound the probability
that $S\in\cantor$ has the property that for the given value of $k$
there is $p$ such that
\begin{equation}\label{eqn:prop}
(\exists t)\ U_t(p)=S\uph k|p|.
\end{equation}
Since there are $2^\ell$ programs of length $\ell$, the probability that there is a program $p$ of length $\ell$ such that \eqref{eqn:prop} holds is at most $2^\ell/2^{k\ell}$. Adding up over all possible lengths $\ell$ we obtain
\begin{equation}
 P_{{\rm error}}\leq \sum_{\ell> 0}\frac{2^\ell}{2^{k\ell}}=\frac{2^{-(k-1)}}{1-2^{-(k-1)}}=O\left(2^{-k}\right),
\end{equation}
which goes to zero exponentailly with $k$. 

The protocol would then work as follows: given a tolerated error
probability $e$, one chooses a $k$ large enough so that the
previous bound is smaller than $e$, and then run the described
algorithm with inputs $k$, $X$ and $Z$.

\subsection{Noise robustness}

We now show how to modify the previous algorithm to make it robust
against noise. We can consider a very natural noise model in which
random bit flips are applied to the measured sequences, resulting
for instance from imperfect preparations or measurements.
Therefore, we modify the algorithm so that it tolerates a fraction
$q \in \mathbb{Q}$ of bit flips in the prefixes. The modified
algorithm is as follows:

\begin{algorithm}\caption{The noise tolerant distinguishing protocol}\label{alg:distinguishNoiseAP}
\begin{algorithmic}
\Require $q\in \mathbb{Q}$, $k \in \NN$ and $X,Z\in\cantor$, one of them being computable
\Ensure `$X$' or `$Z$' as the candidate for being computable; wrong answer with probability bounded by $O(2^{-k})$
\For{$t=0,1,2\dots$}
    \For{$p=0,\dots,t$}
            \If {$d_H(U_t(p),X\uph k|p|)<qk|p|$} \State output `$X$' and halt \EndIf
            \If {$d_H(U_t(p),Z\uph k|p|)<qk|p|$} \State output `$Z$' and halt \EndIf
    \EndFor
\EndFor
\end{algorithmic}
\end{algorithm}
where $d_H$ is the Hamming distance between two strings, which
counts the number of different bits in both strings. The first
thing to notice is that when $q=0$ Algorithms
\ref{alg:distinguishAP} and \ref{alg:distinguishNoiseAP} coincide.

We need to show now that, again, the success probability can be
made as close to one as desired by choosing the parameter $k$ and
that the algorithm always halts. Instead of bounding the number of
sequences that can be generated with a program of length $\ell$,
 we need to bound the
number of sequences that have a Hamming distance smaller than
$qk\ell$ from a computable one. One possible bound is $2^\ell
\binom{\ell k}{\left\lfloor q\ell k \right\rfloor} 2^{\left\lfloor q\ell k
\right\rfloor}$, where the first exponential term counts the
number of different programs of length $\ell$, the combinatorial
number corresponds to the number of bits that can be flipped due
to errors, and the last exponential term gives which of these bits
are actually being flipped. This estimation may not be tight, as
we may be counting the same sequence several times. However, using
this estimation we derive a good enough upper bound on the final
error probability, as we get

\begin{equation}
 P_{{\rm error}}<\sum_{\ell >0}\frac{2^\ell 2^{\left\lfloor q\ell k \right\rfloor}\binom{\ell k}{\left\lfloor q\ell k \right\rfloor}}{2^{\ell k}}
\end{equation}
If we consider that $q<1/2$, we can remove the integer part
function and use the generalization of combinatorial numbers for
real values. Then, by using that
$\binom{a}{b}\leq\left(\frac{ea}{b}\right)^b$, we obtain

\begin{equation}
  P_{{\rm error}}<\sum_{\ell >0}\left[2^{(1+qk-k)}  \left(\frac{e}{q}\right)^{qk}\right]^\ell.
\end{equation}
This geometric sum can be easily computed yielding

\begin{equation}
 P_{{\rm error}}<\frac{2^{1+qk-k}\left(\frac{e}{q}\right)^{qk}}{1-2^{1+qk-k}\left(\frac{e}{q}\right)^{qk}}.
\end{equation}
Now, it can be shown numerically that for $q\lesssim 0.21$ the
probability of mis-recognition tends to zero exponentially with
$k$.

Finally, we need to show that the noise tolerant algorithm always
halts. Let $r<q$ be the probability of a bit flip. By the definition
of probability we now have that for every $\delta$ there exist an
$m_0$ such that for every $m>m_0$ the portion of bit flips in both
$X\uph m$ and $Z\uph m$ are less than $(r+\delta)m$. This
means that if we go to long enough prefixes (or programs), the
portion of bit flips will be less than $q$. And since any
computable sequence is computable by arbitrarily large programs,
this ensures that our algorithm will, at some point, come to an
end.

\subsection{Discussion}


We have shown that if Alice uses a computing device satisfying the
Church-Turing thesis thesis to prepare a seemingly proper mixture
of $\ket{0}$ and $\ket{1}$ or a seemingly proper mixture of
$\ket{+}$ and $\ket{-}$, both apparently yielding the maximally
mixed state, Bob can distinguish both situations.

It is worth noticing that, although our algorithm halts in finite
time, it can take extremely long, depending on the length of the
shortest program that generates the needed prefix and the time it
takes to find it. Nonetheless, what we have shown with this is
that in both preparations, somehow, the resulting state has
information on how it was prepared. Our algorithm can be thought
of as a tool to prove that both preparations are indeed
distinguishable, but there might be protocols that finish in
shorter times. And even if there are not, the fact that both
situations are distinguishable still holds, showing that having a
computable preparation leaves a mark on the states it produces.

This apparent paradox can be easily resolved the following way:
computable sequences have correlations that we are not taking into
account. This means that Alice's choice is not given, as needed,
by a set of independent and identically distributed random
variables but by a computable sequence. The evident consequence of
this is that Bob can distinguish both situations and our main
results is to provide such an algorithm.

It is not clear whether a proper quantum state can be associated
to each qubit leaving Alice's box. Let us imagine a situation in
which Alice has already given Bob several qubits (as many qubits
as Bob wanted to request from Alice), and we ask Bob to guess the
next qubit
---unknown to him---, with the sole
promise that Alice, in the limit, will pick as many states from
one bag as from the other (i.e.\ it is a balanced sequence). Since
every prefix can be extended to a computable sequence, no matter
what Bob already knows about Alice's preparation, he cannot say
anything about the next qubit. For instance, he can already know
in what basis Alice is preparing each state (via the presented
algorithm), and an extremely long prefix of the computable
sequence that Alice is using. Still, he does not know if the next
qubit will be $\ket{0}$ or $\ket{1}$ (if Alice prepares in the
computational basis). The best description for that single qubit
state that Bob can give, from the balanced sequence promise, is
$\rho={\mathbb I}/2$.

Interestingly, our results easily extend to other ensembles, and
can for instance be applied to the mixed states experimentally
produced using a classical random number generator of
\cite{amselem2009experimental,PhysRevLett.105.130501}. Our
classical algorithm is also suitable for performing other
seemingly impossible tasks. If Bob is presented with two states,
one that is a proper computable mixture of the states $\ket{0}$
and $\ket{1}$ each with equal weight, and the other is an improper
mixture yielding the maximally mixed state (for instance, one of
the parts of a maximally entangled state), Bob can distinguish
which is which by a slight modification of our algorithm. He just
obtains two sequences, each by measuring $\sigma_z$ to each state.
Again, the problem is reduced to distinguishing, with high
probability, a fair coin from a computable sequence, a task that
we have already shown how to solve.

\section{The Bell test computability loophole\label{apBell}}

In this section we discuss how the knowledge that measurement
settings are chosen using a device that satisfies the
Church-Turing thesis opens a new loophole in Bell tests. We focus
our attention on the simplest Bell scenario although
generalizations are straightforward: let us consider a bipartite
scenario in which the two parties, Alice and Bob, have a box each
with two input buttons (left and right) and a binary output. A
source between them is sending physical systems sequentially to
each party. Upon arrival, Alice and Bob choose what input buttons
to press thus performing different measurements on the particles.
Our object of interest is the probability distribution describing
the process $P\left(a,b|x,y\right)$ where $x$ and $y$ are inputs
for Alice and Bob's box respectively and $a$ and $b$ are their
outputs which can be derived from the statistics.

Let us also imagine that the input-output events at each site
define space-like separated events so that Bob's input cannot
influence Alice's output and viceversa. Focusing on a particular
round of the experiment, let us describe by $\lambda$ a complete
set of variables (some of which could be hidden or unknown) that
characterize the physical systems such that the outcomes of the
measurements are deterministic
$P(a,b|x,y,\lambda)=\delta_{f(x,\lambda)}^a
\delta_{g{(y,\lambda)}}^b$ where the functions $f(x,\lambda)$ and
$g(y,\lambda)$ map determinstically inputs $x,y$ to outputs $a,b$,
using the complete description $\lambda$. Different rounds of the
experiment may be described with different set of variables
$\lambda$ drawing from a probability distribution
$p(\lambda|x,y)\equiv p(\lambda)$ where the equality condition is
termed {\em measurement independence} or {\em free choice}. This
crucial condition implies that the complete description $\lambda$
is independent from the choice of settings $x,y$ of Alice and Bob
will use to measure the systems. Thus, we  say that a probability
distribution is {\em local} if it can be written as

\begin{equation}
 P\left(a,b|x,y\right)=\int p(\lambda) \delta_{f(x,\lambda)}^a \delta_{g{(y,\lambda)}}^b d\lambda.
\end{equation}

It can be shown that any probability distribution that violates the following Bell inequality, namely a CHSH inequality~\cite{Clauser1969}, is not a local distribution:
\begin{equation}\label{eqCHSH}
\sum_{a,b,x,y\in \{0,1\}} (-1)^{a+b+x\cdot y} P(a,b|x,y) \leq 2.
\end{equation}
Remarkably, quantum correlations, obtained for example by
measuring a maximally entangled two-qubit state with non-commuting
measurements, can violate this inequality. The violation of a Bell
inequality witnesses the existence of non-local correlations which
in turn can be used in many device-independent applications such
as randomness expansion or for establishing a secure key between
distant locations. Hence, checking whether the experimental data
truly violates a Bell Inequality is of outmost importance for
device-independent information science.

\begin{figure}[h]
\includegraphics[width=8cm]{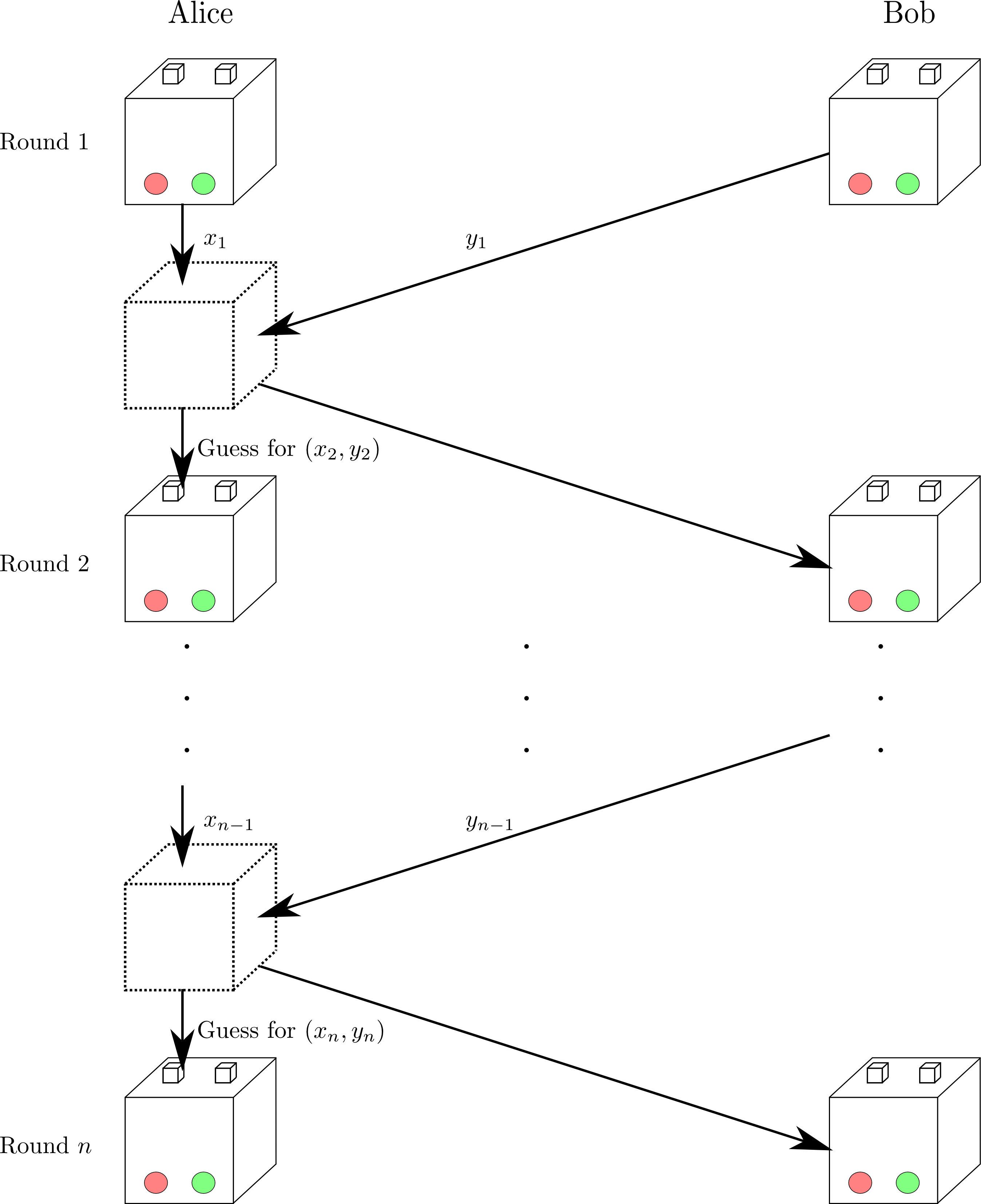}
\caption{Scheme for the Bell inequality computability loophole. After each round $i$, Alice's box receives the information about  Bob's choice of measurement $y_i$. Using the information from all previous choices of inputs for both parties, Alice's box makes a prediction on what the next round inputs will be by using the presented algorithm. For simplicity in the notation, the attack is shown for a Bell test involving two measurements per party. As long as Alice, Bob or both use a computable sequence for their choice, the guess will start being correct after a number of rounds. Once this happens, the boxes can simulate any probability distribution, both local and non-local. Therefore Alice and Bob can not rule out an eavesdropper having prepared their boxes.\label{figAliceBobAP}}
\end{figure}

In order to present the computability loophole, we introduce an
eavesdropper named Eve. Eve will be able to prepare Alice's and
Bob's local boxes at the beginning of the experiment as shown in
Fig.~\ref{figAliceBobAP}. Alice's box will then have access to
the inputs of both parties after the measurements are done
(Alice's input in a straightforward way, and Bob's input via
classical communication). This scenario was first termed the
two-sided memory loophole in the literature
\cite{PhysRevA.66.042111}. An equivalent scenario has been used
more recently in \cite{pironio2010random,PhysRevA.87.012336} so as
to perform device-independent randomness expansion where they
allow the boxes behaviour to adapt depending on the information of
previous rounds. Interestingly, local models exploiting this past
information have been shown to be of no help to violate the Bell
Inequality in the asymptotic limit. Indeed, the probability that a
local model reproduces some observed violation despite using past
inputs and outputs goes exponentially fast to zero in the number
of rounds \cite{PhysRevA.87.012336}.

As mentioned earlier, a crucial condition for Bell tests to
establish nonlocality and randomness is to assume measurement
independence $p(\lambda |x,y)\equiv p(\lambda)$, which in our
context is independence between the boxes prepared by Eve and the
measurement choices of Alice and Bob. Let us imagine that Alice
and Bob chose their measurement settings following an algorithm
which is a standard practice in all Bell experiments to date.
Trivially, if Alice's box knows which algorithms Alice and Bob are
using, she can fake a Bell violation. We assume that the
algorithms used by Alice and Bob are fully unknown to each other
and to Eve, thus uncorrelated to the boxes she initially prepared.

Our result is to show that if either Alice or Bob (or both) choose
their measurements following an algorithm ---or equivalently,
because of the Church-Turing thesis, following any classical
mechanical procedure---, even under the assumption that such
algorithm is fully unknown to Eve and hence uncorrelated to the
boxes she initially prepares, there is an attack that, in the
asymptotic limit, produces a Bell inequality violation between
Alice and from purely deterministic boxes, thus providing the
aforementioned loophole.

Before proceeding, we need a few more tools from computer science.
We say that a class $\CC$ of computable functions
is a {\em time [resp. space] complexity class} if there is a
computable function $t$ such that each function in $\CC$ is
computed by a Turing machine that, for every input $x$, runs in
time [resp. space] $O(t(|x|))$. Examples of such classes include
the well-known {\bf P},  {\bf BQP}, {\bf NP}, {\bf PSPACE} where
the complexity time bound is a simple exponential function and the
much broader class {\bf PR} of the primitive recursive functions
where the time bound is Ackermannian
\cite[\S{VIII.8}]{Odifreddi1999} (see \cite{complexityZoo} for an
inclusion diagram of the most well-studied complexity classes).


Let us now say that one party, say Alice without loss of
generality, will be using an algorithm to produce her measurements
choices. In formal terms, this means that there is a computable
function $f_A:\NN\to\{{\rm left},{\rm right}\}$ such that $f_A(i)$
tells Alice to press the left or the right button at the $i$-th
round.

As we previously pointed out, it is clear that if Eve knows (any
algorithm for) $f_A$, her task becomes trivial. In our setting,
however, function $f_A$ is unknown to Eve when she prepared the
boxes. However, we assume the following further hypothesis: Eve
knows {\em some} time or space bound $t$ of a complexity class
containing $f_A$ and $f_B$ (the corresponding function for Bob's inputs). For instance, Eve knows that Alice and
Bob use at most, say, time $O(t(n))$, for $t(n)=2^{2^n}$
(though the algorithms that Alice is actually running may take,
say, $O(n^2)$). It is important to note that this hypothesis is
quite mild, because every computable function belongs to some time
or space complexity class ---given a program there is a computable
interpreter which executes it on some given input by stages and
counts the number of steps that such execution takes to terminate or the number
of cells used in the tape.
In other words, for every computable function $g$ there is a
computable function $t_g$ that upper bounds the running time or
space of some algorithm for $g$.

Knowing this time or space bound $t$, Eve can program a computing
device in one of the boxes, say Alice's, to {\em predict} the
functions $f_A$ and $f_B$ from some point onwards. This means that
Alice's box has an effective procedure that, after having seen
$f_A(0),f_A(1),\dots, f_A(k)$ for large enough $k$, allows her to
correctly guess $f_A(k+1), f_A(k+2)\dots$ and the same for $f_B$.
The existence of such $k$ will be guaranteed by Alice's box
procedure; however it will not be able to effectively determine
when this $k$ has arrived.
The idea behind this is that every time bounded class is
computably enumerable, allowing Eve to pick, every time, the first
program for a function from that class that reproduces the inputs given by Alice
(or Bob) so far. Since the function used by Alice (or Bob)
belongs to that class, at some point the first program that Alice
will find reproducing the inputs given so far will be one which computes the function
used by Alice (or Bob), therefore allowing Alice's box to
predict every input to come. See Sec. \ref{sec:learnability} for a
detail of Alice's box procedure.

Back to the loophole, under these assumptions, Eve is able to
prepare both boxes so as to fake a Bell Inequality violation.
Moreover, she could even prepare boxes that seem to be more
non-local than what quantum mechanics allows. To see how, notice
that any no-signaling bipartite probability distribution, local or
not, can always be written as
\begin{eqnarray}
P(a,b|x,y)&=&\int p(\lambda) \delta_{f(x,\lambda)}^a \delta_{g(y,x,\lambda)}^b d\lambda\\
&=&\int p'(\lambda) \delta_{f'(y,\lambda)}^b \delta_{g'(x,y,\lambda)}^a d\lambda
\end{eqnarray}
where again functions $f,f',g,g'$ are deterministic functions (See
Sec. \ref{sec:nscorr} for a prove). This means that, given that
Eve learns either Alice's input $x$ or Bob's input $y$, she can
prepare deterministic (local) boxes to simulate any probability
distribution and hence fake any Bell Inequality violation.

\subsection{Predicting computable functions from initial segments}\label{sec:learnability}

The theory of predicting computable functions started with the
seminal works by Solomonoff
\cite{solomonoff1964formalP1,solomonoff1964formalP2} on inductive
inference, and Gold \cite{gold1967language} on learnability. It
studies the process of coming up with, either explanations (in the
form of computer programs) or next-value predictions, after seeing
some sufficiently big subset of the graph of a computable
function. Many possible formalizations, depending on how the data
is presented and how the learning process converges, have been
considered in the literature (see \cite{zeugmann2008learning} for
a comprehensive survey). The most suitable model for our purposes
is called {\em identification by next value}, and follows by
elementary arguments from computability theory.
%

A class  of total computable functions $\Class$ is
\emph{identifiable by next value} ($\Class \in \NV$)
\cite{barzdin1971prognostication} if there exists a computable
function $g$ (called a {\em next-value function for $\Class$})
such that for every $f \in \Class$,
\begin{equation}\label{eqn:predict}
(\exists n_0)(\forall n\geq n_0)\ f(n) = g(\langle f(0),\dots,f(n-1)\rangle).
\end{equation}
Here $\langle x_1,\dots,x_n\rangle$ is any computable codification
of an $n$-tuple with a natural number, whose decoding is also
computable. Condition \eqref{eqn:predict} formalizes the idea that
given the past values of $f$ (namely $\left(f(0),\dots,f(n-1)\right)$), $g$ can
predict the forthcoming value of $f$ (namely, $f(n)$), provided
$n$ is large enough ---how large depends on the function $f$ that
we want to learn.

%
%
%

It follows from a simple diagonal argument that the class of all
computable functions is not in $\NV$. However, any time or space
complexity class is $\NV$. Indeed, suppose $\Class$ is a time
complexity class with (computable) time bound $t$. The following
algorithm computes a next-value function for $\Class$. Let
$(M_i)_{i\in\NN}$ be an enumeration of all Turing machines.

\begin{algorithm}\caption{A next-value algorithm for a time class $\Class$ with bound $t$}\label{alg:nextvalueAP}
\begin{algorithmic}
    \Require $n \in N$
    \Ensure $g(n)$, the next-value function for $\Class$.
    \State Let $n=\langle m_0,\dots, m_{n-1}\rangle$
    \State Let $\langle e,c\rangle$ be the least number such that
        \begin{enumerate}
        \item[i.] for $i\in\{0,\dots,n\}$, $M_e(i)$ halts after at most $c\cdot t(|i|)$
        \item[ii.] for $i\in\{0,\dots,n-1\}$ $M_e(n)$ outputs $m_i$
        \end{enumerate}
    \State Output $M_e(n)$
\end{algorithmic}
\end{algorithm}

Suppose $f\in\Class$, i.e.\ there is some Turing machine $M_{e'}$ and constant $c'$ such that for every $x\in\NN$,
\begin{equation}\label{eqn:true-ec}
\mbox{$M_{e'}(x)$ computes $f(x)$ with time bound $c'\cdot t(|x|)$.}
\end{equation}
 Both $e'$ and $c'$ are unknown, and the idea of  Algorithm \ref{alg:nextvalueAP} is to try different candidates $e$ and $c$ for $e'$ and $c'$ respectively, until one is found.
On input $n=\langle f(0),\dots,f(n-1)\rangle$ the algorithm proposes a {\em candidate} Turing machine $M_e$ which `looks like' $f$ on $0,\dots,n-1$, and then guesses that $f(n)$ is the value computed by $M_e(n)$. To be a candidate means not only to compute the same first $n$ values, but also to do it within the time bound imposed by $\Class$, which is $c\cdot t$.
Of course, the chosen candidate may be incorrect because, for instance, on input $\langle f(0),\dots,f(n-1),f(n)\rangle$ we may realize that $f(n)$ was not equal to $M_e(n)$. In this case, the algorithm changes its mind and proposes as candidate a new pair $\langle e,c\rangle$. The existence of the correct candidates $e'$ and $c'$ satisfying \eqref{eqn:true-ec} guarantees that:
\begin{enumerate}
\item For each $n$ and each input $\langle f(0),\dots,f(n-1)\rangle$ the algorithm will find some $\langle e,c\rangle$ meeting conditions i and ii.
\item Along the initial segments $\langle f(0),\dots,f(n-1)\rangle$ for larger and larger $n$ there can only be finitely many mind changes. Indeed, if the number of mind changes were infinite, then $\langle e',c'\rangle$ would be ruled out and this is impossible, as conditions i and ii are true for $e=e'$ and $c=c'$.
\end{enumerate}

Hence there is $n_0$ such that for all $n\geq n_0$, the algorithm makes no more mind changes, and it stabilizes with values $\langle e,c\rangle$, which may not necessarily be equal to $\langle e',c'\rangle$, but will satisfy that $M_e(x)$ computes $f(x)$ with time bound $c\cdot t(|x|)$. Thus on input $\langle f(0),\dots,f(n-1)\rangle$ the algorithm will return $f(n)$, and hence \eqref{eqn:predict} will be satisfied. Observe that although the algorithm starts correctly predicting $f$ from one point onwards, it cannot detect when this begin to happen. In other words, $n_0$ is not uniformly computable from $e'$ and $c'$.

The algorithm for a space complexity class with bound $t$ is analogous, but condition i must be modified to
\begin{enumerate}
\item[i'.] for $i\in\{0,\dots,n\}$, $M_e(i)$ halts after at most $2^{t(|i|)}$ many steps and uses at most $t(|i|)$ many cells of the work tape during its computation.
\end{enumerate}
Observe that any halting computation which consumes $t(n)$ many cells of the work tape runs for at most $2^{t(n)}$ many steps, as this is the total number of possible memory configurations. In condition i' we add the statement on the number of steps in order to avoid those computations which use at most $t(n)$ many cells but are non-terminating.

\subsubsection{Simulation of no-signaling correlations from deterministic boxes}\label{sec:nscorr}

For completeness, we give a simple proof of the well-known fact
that, if the input of one party in a Bell test is known, one can
simulate any no-signaling distribution by using deterministic
boxes. Let us imagine without loss of generality that it is Bob's
box the one that has access to Alice's input $x$. First, notice
that any no-signaling box can be written in the following way

\begin{equation}
P(a,b|x,y)=P(a|x)P(b|y;a,x)\equiv P(a|x)P_{a,x}(b|y). \\
\end{equation}

Trivially, any local distribution can be simulated through
deterministic boxes as $P(a|x)=\int p(\lambda)
\delta_{f(x,\lambda)}^a d\lambda$. Hence, the no-signaling
bipartite distribution can be written as

\begin{equation}
P(a,b|x,y)=\int \int p(\lambda)p'(\lambda') \delta_{f(x,\lambda)}^a \delta_{g(y,\lambda',a,x)}^b d\lambda d\lambda'
\end{equation}
by defining now $\lambda''=(\lambda, \lambda')$ and therefore
$d\lambda''=d\lambda d\lambda'$ and
$p'(\lambda'')=p(\lambda)p'(\lambda')$ we have

\begin{equation}
P(a,b|x,y)=\int p''(\lambda)\delta_{f(x,\lambda'')}^a \delta_{g(y,\lambda'',x)}^b d\lambda''.
\end{equation}

Notice that since $a$ is a deterministic function of $x$ and
$\lambda$, given that $\lambda''$ includes the information of
$\lambda$, the function on Bob's side does not need to depend
explicitly on $a$.

\subsection{Discussion}

We have shown that if either Alice or Bob choose their inputs for
a Bell experiment in a computable way, an eavesdropper
restricted to prepare deterministic devices can make them believe
to have non-local boxes, thus creating a computability loophole.
Notice that this scenario is equivalent to letting the boxes
communicate before the runs and adapt accordingly, as is the case
in the randomness expansion protocols
\cite{PhysRevA.87.012336,pironio2010random} where our loophole
would also apply if either Alice or Bob would use
pseudo-randomness. There is no way of preventing this form of
communication, unless some assumptions on shielding of the devices
are enforced.

From a fundamental perspective, our result answers the question about what type of randomness is necessary for having a valid violation of a Bell inequality: 
Alice and Bob's behaviour need to be non-algorithmic. 
Therefore no computable pseudo randomness criterion will suffice for a proper Bell inequality violation. It is natural to ask, at this point, where can Alice and Bob find sources of true randomness for their inputs.
If they assume quantum mechanics, then flipping a quantum coin would suffice (with probability one). 
However, it is not desirable to assume a non-local theory like
quantum mechanics, in order to test non-locality. Ruling out
quantum mechanics, the other source of true randomness that is
usually mentioned is free will. However, there is no present
evidence that the human brain is able to produce non-algorithmic
randomness.




\bibliography{thesis_bib}
\bibliographystyle{unsrt}

\end{document}